# Magnetic support of the optical emission line filaments in NGC 1275

A. C. Fabian[1], R. M. Johnstone[1], J. S. Sanders[1], C. J. Conselice[2], C. S. Crawford[1],
J. S. Gallagher III[3] & E. Zweibel[3,4]

[1]Institute of Astronomy, Madingley Road, Cambridge CB3 0HA, UK. [2]University of Nottingham, School of Physics & Astronomy, Nottingham NG7 2RD, UK. [3]Department of Astronomy, University of Wisconsin, Madison, WI 53706, USA. [4]Department of Physics, University of Wisconsin, Madison, WI 53706, USA.

**The giant elliptical galaxy NGC 1275, at the centre of the Perseus cluster, is surrounded by a well-known giant nebulosity of emission-line filaments[1,2], which are plausibly about >$10^8$ yr old[3]. The filaments are dragged out from the centre of the galaxy by the radio bubbles rising buoyantly in the hot intracluster gas[4] before later falling back. They act as dramatic markers of the feedback process by which energy is transferred from the central massive black hole to the surrounding gas. The mechanism by which the filaments are stabilized against tidal shear and dissipation into the surrounding $4\times10^7$ K gas has been unclear. Here we report new observations that resolve thread-like structures in the filaments. Some threads extend over 6 kpc, yet are only 70 pc wide. We conclude that magnetic fields in the threads, in pressure balance with the surrounding gas, stabilize the filaments, so allowing a large mass of cold gas to accumulate and delay star formation.**

The images presented here (Figs. 1—4) were taken with the Advanced Camera for Surveys (ACS) on the NASA Hubble Space Telescope (HST) using three filters; F625W in the red contains the Hα line, F550M is mostly continuum and F435W in the blue which highlights young stars. In Fig. 2 we show part of the Northern filament ~27 kpc from the nucleus (we adopt $H_0$=71 km s$^{-1}$ Mpc$^{-1}$ which at a redshift 0.0176 for NGC 1275 gives 352 pc arcsec$^{-1}$). The filaments seen in the WIYN ground-based image (right) are just resolved into narrow threads with the HST ACS (see Supplementary Information). This also occurs in many other filaments including the north-west "horseshoe" filament (Fig. 3) which lies immediately interior to the outer ghost bubble in X-ray images[5]. A fine thread of emission is seen in the Northern filament system extending about 16 arcsec or 5.8 kpc. Averaged over kpc strips it is about 4 pixels (0.2 arcsec) or about 70 pc wide. (This is an upper limit as the point spread function of the ACS is about one half this value.) The aspect ratio (length / thickness) therefore approaches 100. The top of the horseshoe which is about 6 kpc across is similar, as are many other relatively isolated filaments.

In order to estimate the required magnetic field we need to know the properties of a filament and its surroundings. We shall concentrate on a thread of radius 35 pc and length 6 kpc at a distance of 25 kpc from the nucleus of NGC 1275 (Fig. 2) as a basic structural unit typical of what is now resolved in the filaments. To estimate the mass for such a thread we scale from the total gas mass of $10^8$ M$_\odot$ inferred from CO emission[6] observed in a 22 arcsec IRAM beam on the same Northern filament complex. Assuming that the mass scales with Hα emission, which is the case for the H$_2$ emission measured with Spitzer[7], then our fiducial thread has a mass of about $10^6$ M$_\odot$. Its mean density is then ~2 cm$^{-3}$ and perpendicular column density $N \sim 4\times10^{20}$ cm$^{-2}$ or $\Sigma_\perp \sim 7\times10^{-4}$ g cm$^{-2}$. The lengthwise column density, $\Sigma_\parallel$, is $l/2r$ times larger.

The variation in projected radial velocity along the filaments is about 100—200 km s$^{-1}$ (ref. 4). If we assume that after correction for projection the velocity shear is about 300 km s$^{-1}$ then, considering the whole structure out to a radius of 50 kpc, it must be 1—$2\times10^8$ yr old. Individual filaments may be in ballistic motion, falling back in for example, but to retain their structure over this time means that something must balance gravity or at least tidal gravitational forces. For a filament of radial length $l$ at galaxy radius $R$ the gravitational acceleration $g \sim v^2/R$, where $v$ is the velocity dispersion in the potential (about 700 km s$^{-1}$ at that radius as inferred from the X-ray measured temperature of the intracluster medium[8]) at radius $R \sim 25$ kpc; tidal acceleration is smaller by $2l/R$. The most likely force to balance a filament against gravity is that due to the tangential component of a magnetic field, as suggested by the filamentary morphology.

Consider first a horizontal thread supported against gravity, the field $B_h$ needed for support is then $B_h \sim (4\pi \Sigma_\perp g)^{1/2}$. For the above values this corresponds to $B_h \sim 24$ µG which is less than the equipartition value of 100 µG so energetically possible. By equipartition we mean that $B^2/8\pi = nkT$ and use the total particle density $n = 0.06$ cm$^{-3}$ and temperature $kT = 4$ keV for the X-ray emitting surrounding gas[8]. For a radial filament the value, $B_r$ is $\Sigma_\parallel^{1/2} \sim 10$ times larger than $B_h$ which exceeds equipartition. If the filaments are falling, then $g$ should be replaced by the tidal acceleration, which is smaller by $l/R$, reducing $B_r$ by a factor of 2, comparable with the equipartition value. Note that the required magnetic field is inversely proportional to the radius of the filamentary threads which means that the high spatial resolution of the HST is essential for demonstrating the high magnetic field required in the NGC 1275 system.

Assuming that the total pressure in the filaments equals the outer thermal pressure means that the filaments are magnetically-dominated (ratio of thermal to magnetic pressure $\beta<1$) and are essentially magnetic molecular structures, similar (but at much higher surrounding pressures) to Galactic molecular clouds[9,10]. Horizontal support would be by magnetic cradles similar to what holds up solar prominences[11]. Vertical support to prevent gas flowing down radial filaments requires an unseen component of horizontal field in the filaments. Small transverse components would be compressed by vertical downflow until their pressure is adequate for vertical support. We assume of course that the magnetic field is coupled to the (low) ionized fraction of the cold gas and that the slowness of ambipolar diffusion links the molecular, atomic and ionized components. The emission from the filaments is well reproduced by an internal heating model, possibly energised by magnetic waves and nonthermal particles powered by the kinetic energy of the clouds[12]. The observed full-width-half-maximum velocity dispersion of ~100 km s$^{-1}$ within a filament[4] may be direct evidence of such waves as indeed expected for Alfvenic turbulence, where half the internal pressure is kinetic and half magnetic. A large field in filaments is also one interpretation of the high Faraday rotation measure seen against the tip of the jet at the centre of NGC 1275[13].

The strong fields implied for their support can stop or delay star formation in the filaments since matter will not be able to collect along field lines. Many massive star clusters were discovered across the face of NGC 1275 in early HST imaging[14,15], most of which do not correlate in position with any filaments. However to the south-east and south there are examples of ordered chains of young star clusters (Fig. 4). They are offset from the nearest Hα filaments by a few kpc. The ones to the SE resemble short streamers and are probably unbound clumps falling apart in the tidal cluster field. This gives an age for the clumps of about 20 Myr. Unless due to chance projection, their proximity with the filaments, particularly the one to the south, shows that stars sometimes do form in the filaments and that at least part of the enormous star cluster system of NGC 1275 originates in this way. In general though there are no obvious star clusters associated with the filaments, so at any given time the star formation rate in any filament must be low; cooled gas does not automatically and rapidly form stars.

The critical surface density $\Sigma_c$ for gravitational instability corresponding to the magnetic field we infer in the filaments is given by $\Sigma_c = B/2\pi\sqrt{G} = 0.062$ ($B$/100µG) g cm$^{-2}$ or a critical column density of $N_c = 4\times10^{22}$ cm$^{-2}$ (assuming hydrogen). Given the mean density of a thread of ~2 cm$^{-3}$ and transverse column density ~$2\times10^{20}$ cm$^{-2}$, they are gravitationally stable. Instability could occur however, if either much of the molecular mass is concentrated in a much smaller thickness (~ 0.1 pc for 50 K, assuming thermal pressure balance) and $\beta>1$, or if the field becomes well ordered parallel to a radial filament axis. Thus gravitational stability depends on the configuration of the molecular and magnetic components.

If a filament acts as a coherent unit due to the magnetic field, then it will interact dynamically with the hot, low density, surrounding gas over a length scale corresponding to its own column density. The above mean value of $4\times10^{20}$ cm$^{-2}$ corresponds to about 2 kpc (~ 6 arcsec) in the surrounding hot gas, so the relative straightness of many filaments over scales of 3—10 kpc or more shows that motions in that gas must be reasonably ordered and not highly turbulent on those scales[5]. We note that the volume-filling component of a filament is probably the 5 million K, soft X-ray emitting phase seen in Chandra X-ray images[16].

Filamentary emission-line structures are common in massive young galaxies[17]. The filament system of NGC 1275 may be a nearby example which can be observed in unprecedented detail. The outer filaments of NGC 1275 present us with magnetically-dominated molecular clouds stretched out for individual inspection, and are the only direct way of seeing the action of the active galactic nucleus power on the surrounding intracluster gas apart from X-ray imaging.

Supplementary Information is linked to the online version of the paper at www.nature.com/nature.

# Acknowledgements

A.C.F. thanks the Royal Society for support.

# Author information

# Figures

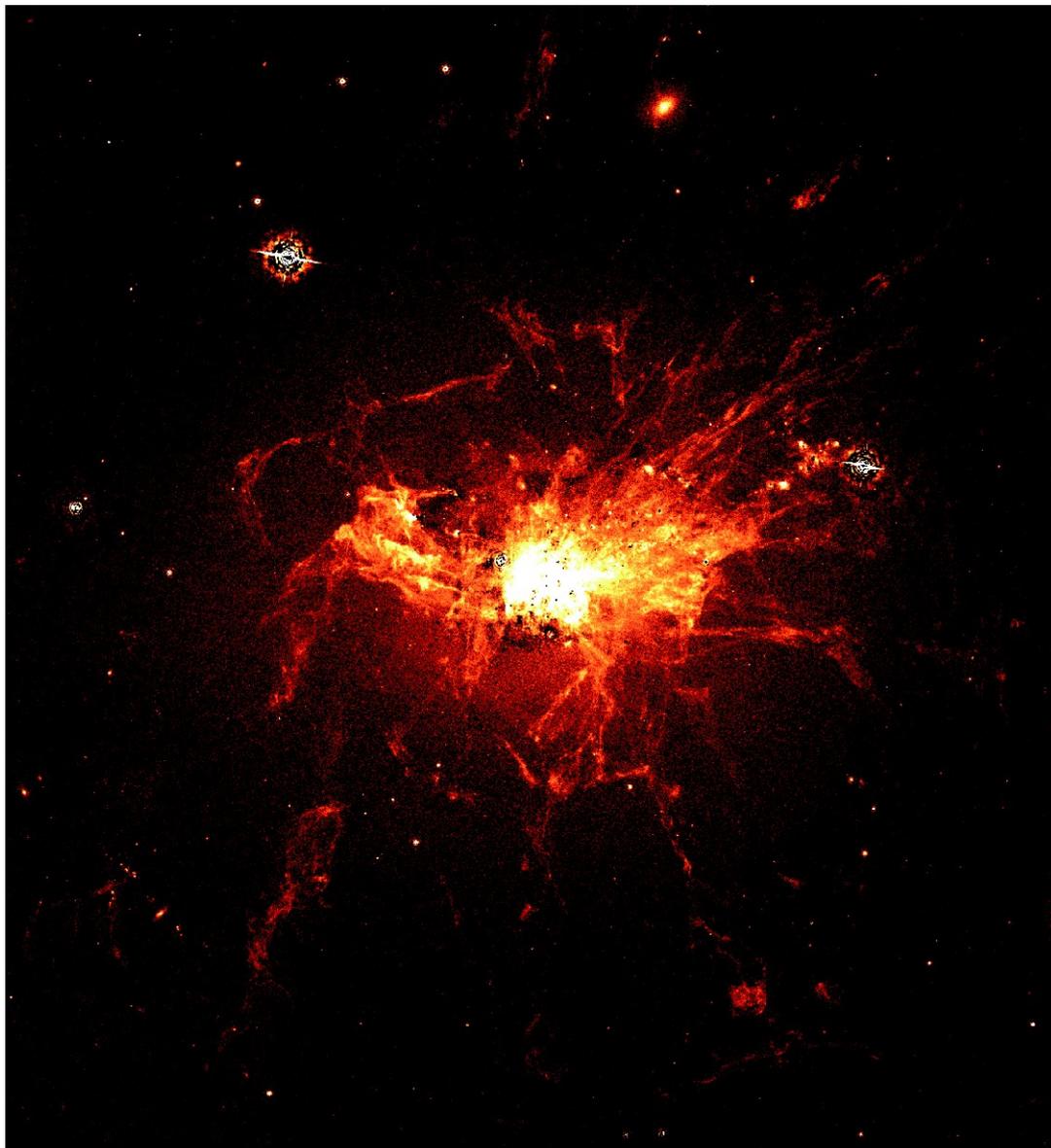

1. Image of the Hα emission from the core of the cluster. This was created by subtracting a scaled green image from the red image, removing the smooth galactic continuum contribution. The image measures 140×150 arcsec in size. Multiple exposures using a three-point line dither pattern were taken at a set of three pointing positions around the centre of NGC 1275 in filters F550M and F625W with considerable overlap between the pointings. Similar data were obtained in the F435W filter, but from only two pointing positions because of the failure of ACS before the completion of this programme. The data from each filter were registered relative to each other using stars and then combined separately into large mosaic images using the latest version of the stsdas task multidrizzle[18].

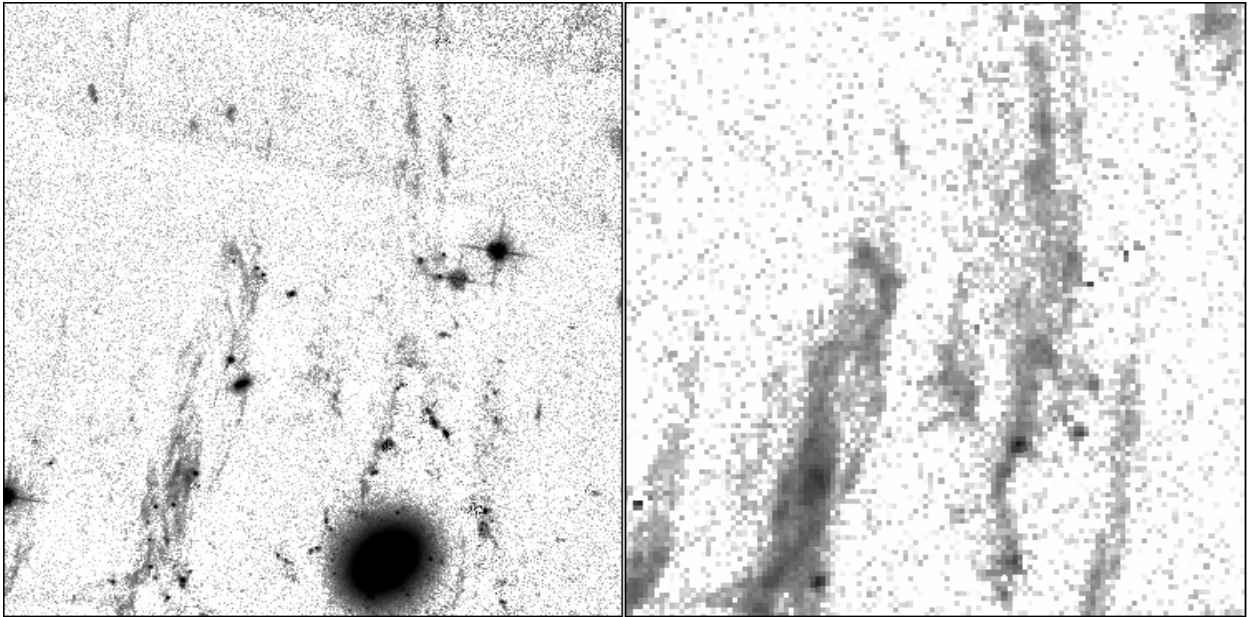

2. Comparison of our new image of the Northern filament system (about 25 kpc north of the nucleus) with that from the WIYN telescope[2]. The ACS image was produced from the red filter. The SExtractor tool[19] was applied to the data to identify sources and create a model for the smooth galactic light. This was subtracted from the red filter image to enhance the filament. The SExtractor neural network was used to identify stellar sources. These sources were hidden by filling their regions with random values selected from the surrounding pixels. Each image measures 46.6×46.1 arcsec in size.

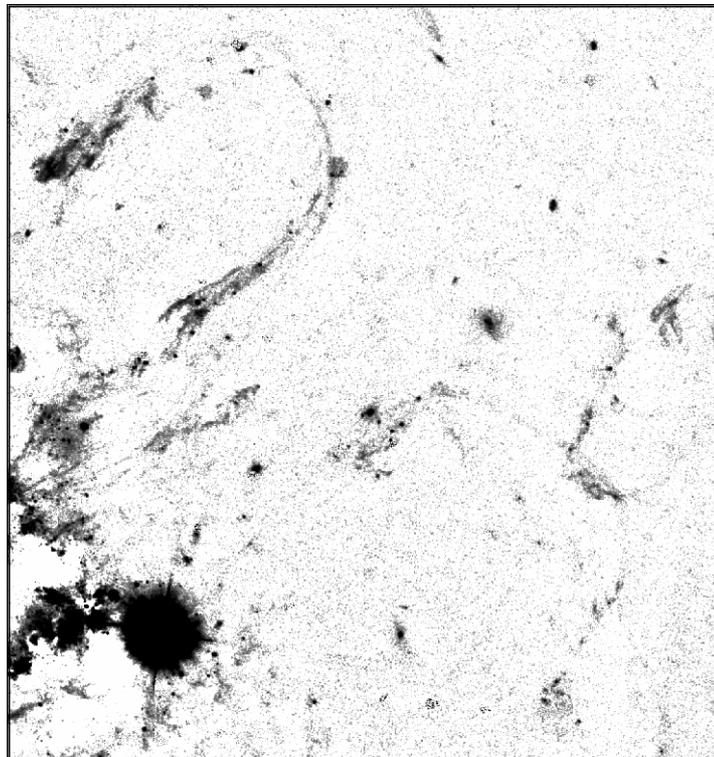

3. Horseshoe filament system about 25 kpc to the north-west of the nucleus. This image has had continuum and stellar sources removed in the same way as Fig 2. The image measures 53.5×56.5 arcsec in size.

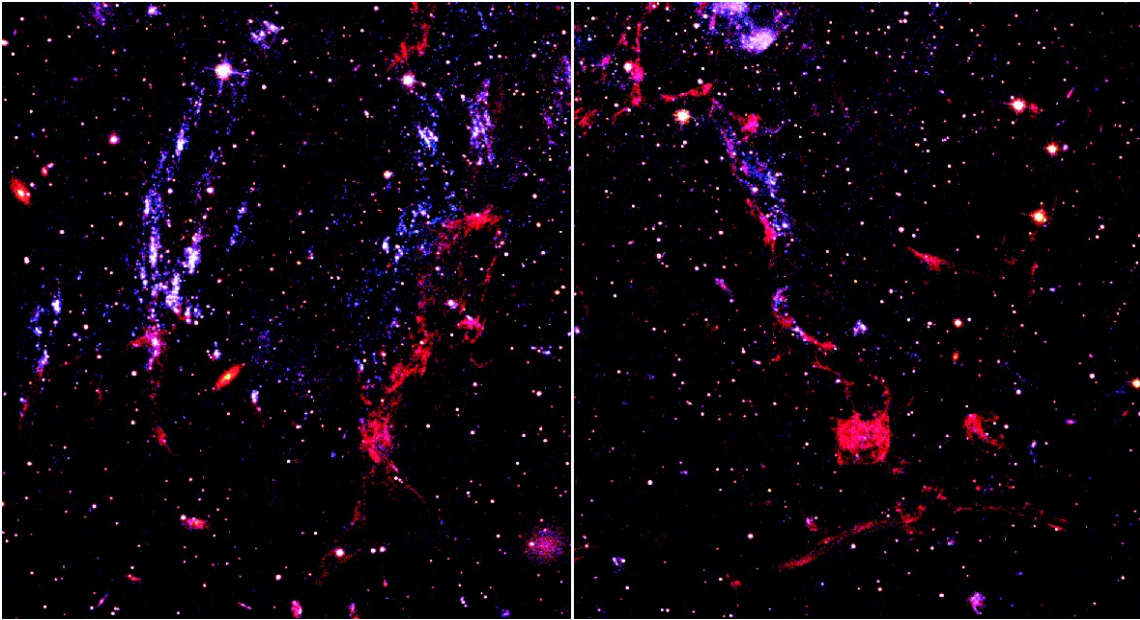

4. Clumps of stars and Hα filaments seen to the south-east and south of the nucleus. SExtractor was used to model the smooth galactic light in the red, green and blue bands. This continuum was subtracted from each image. The images were combined as red, green and blue channels with SAOImage DS9[20].

# Supplementary Notes

We show a complete red, green and blue image in Supplementary Figure 1 and the Hα emission contrasted with the blue emission (mainly young stars) in Supplementary Figure 2. Part of the blue stellar emission to the N of the nucleus is from the High Velocity System (HVS) which is superimposed on NGC 1275, yet must lie 100 kpc or more closer to us (Sanders & Fabian 2007). The "spray" of stars to the north is also likely due to the HVS.

In Supplementary Figure 3 the Hα emission is overlaid on a 2-8 keV Chandra X-ray image. This X-ray band emphasises the pressure of the hot intracluster gas and thus shows the weak shock surrounding the radio bubbles north and south of the centre. It is interesting that the tangential filaments which lie at the ends of several radial filaments (e.g. to the south and west) occur close to this shock.

Closeup images of several features are shown in Supplementary Figure 4. To the left is a "fossil fish"-like structure which straddles the weak shock to the SE. The "tail" of the "fish" has fine structure. The south and east tangential filaments shown in the centre and right images consist of several parallel fine structures.

To assess the thickness of the fine threads we have examined cuts through several of them (Supplementary Figure 5). The images have been drizzled onto 0.025 arcsec pixels so 40 pixels correspond to one arcsec. The profiles left to right correspond to the boxes left to right with the profile of a typical star overlaid in red. The stellar profile is obviously narrower than any of the filaments.

**Supplementary Figure 1**

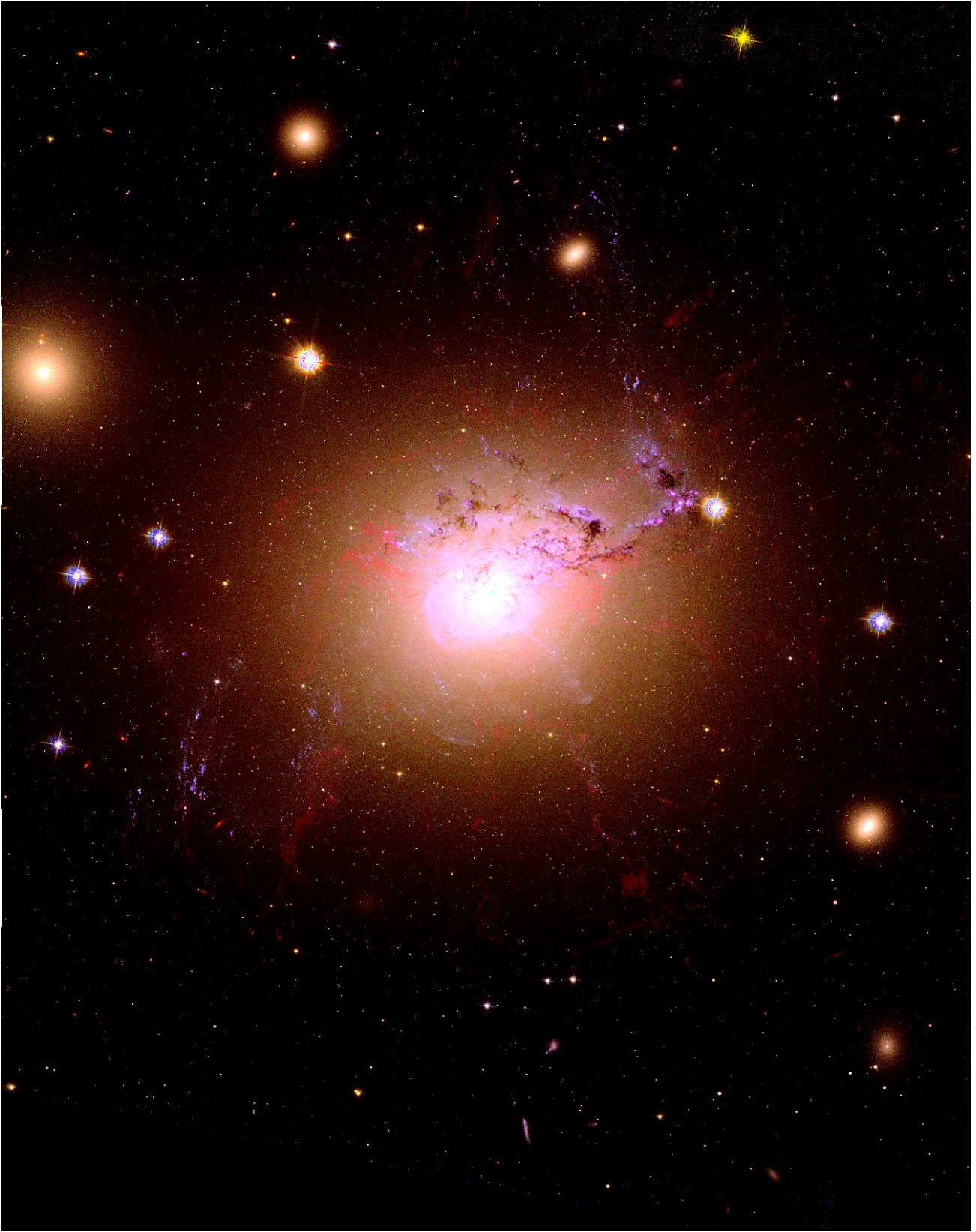

Colour image created by combining the data using the three ACS filters. The three images were processed with the method of Lupton et al (2004) to preserve the colour of objects avoiding saturation. The detail in the Hα filaments was enhanced by using the unsharp mask filter in the GNU Image Manipulation Tool.

**Supplementary Figure 2**

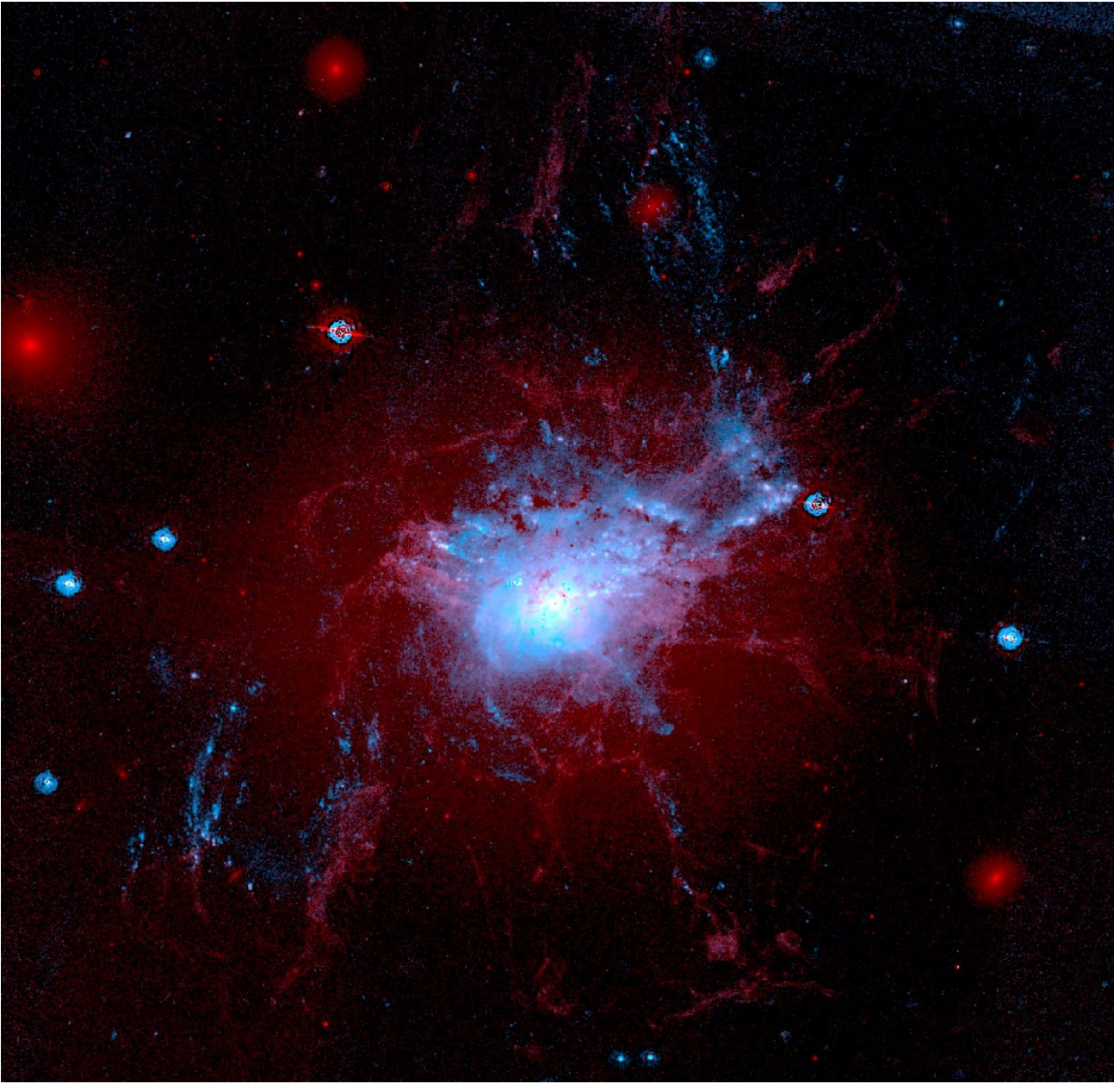

Comparison of the Hα filaments (in red) with the distribution of blue light (in blue). The galactic continuum in the two images was removed by subtracting scaled green images.

**Supplementary Figure 3**

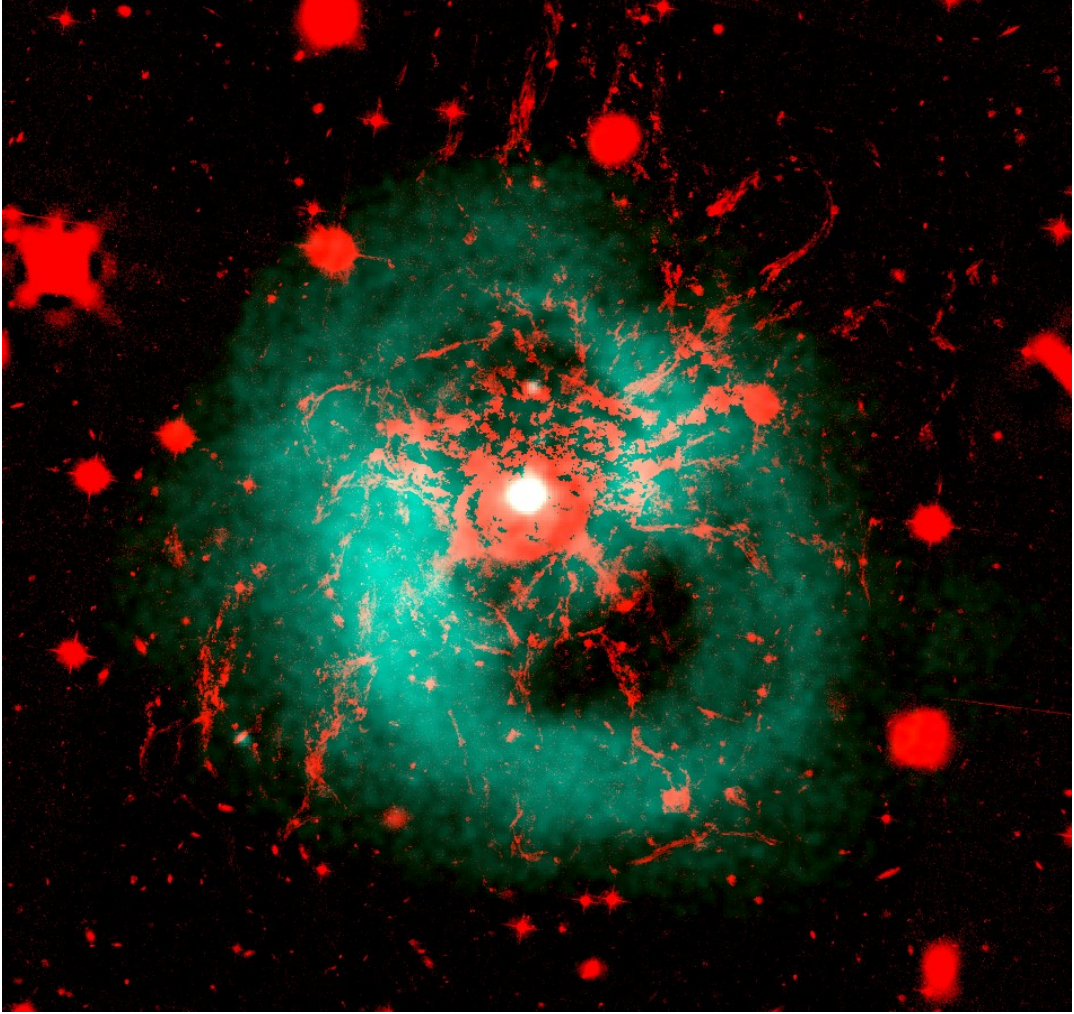

Comparison of the Hα filaments with the hard X-ray emission from the intracluster medium. The Hα image was formed from the red image, subtracting continuum and stellar-sources using SExtractor (as in Fig 2 in the main paper). The X-ray image is 900ks of Chandra data (Fabian et al 2006) in the 2 to 7 keV band, smoothed with a Gaussian with σ=1.5 arcsec.

**Supplementary Figure 4**

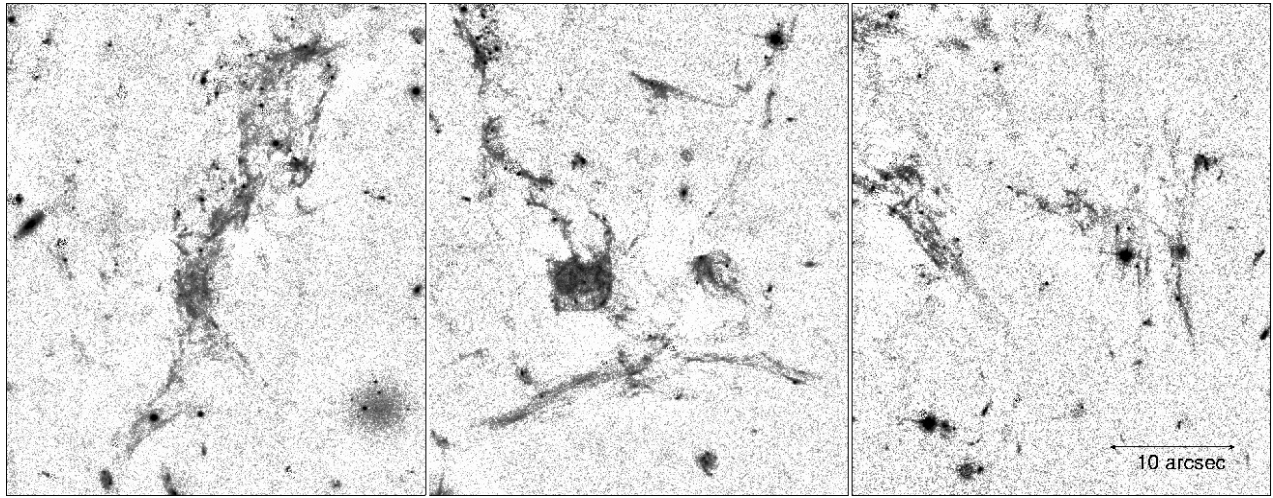

Detail of the `Fish' and the south and west tangential filamentary regions. These are parts of the red image, showing the Hα filaments after subtracting the galactic continuum and stellar-sources using SExtractor (as in Fig 2 in the main paper).

**Supplementary Figure 5**

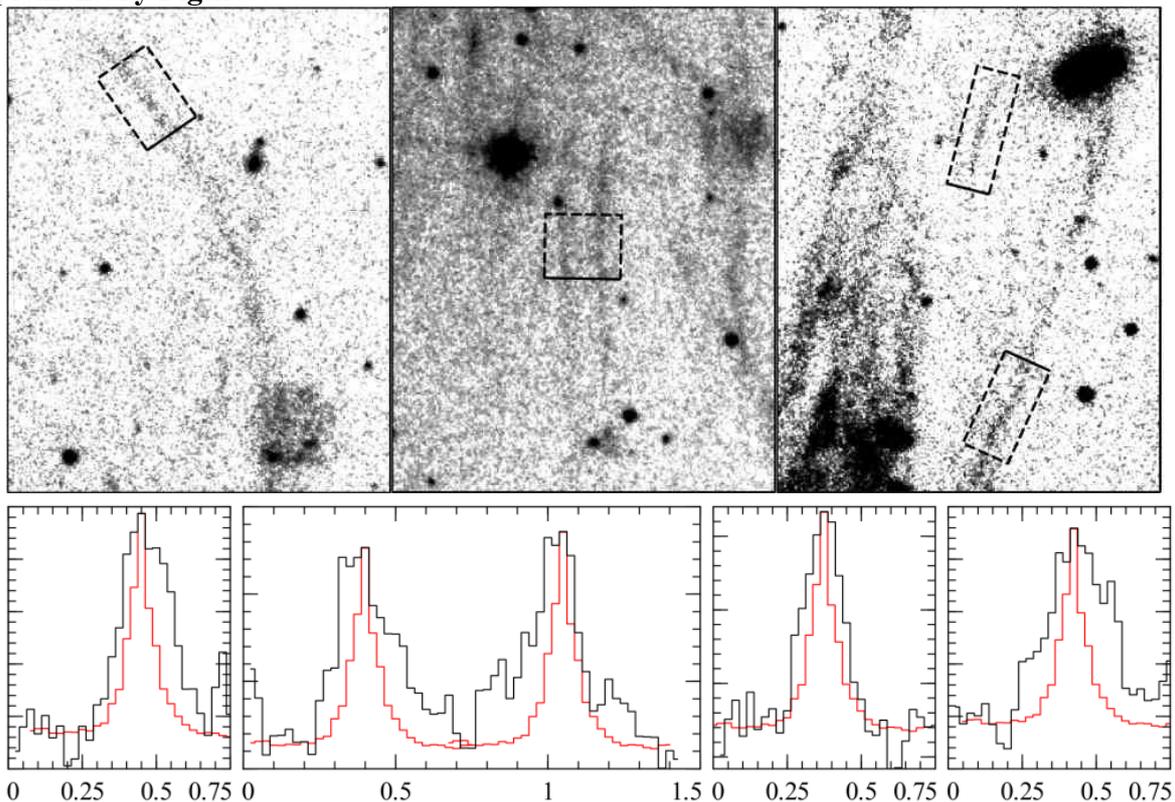

1-D Hα flux profiles of the sections of the filaments marked by boxes. The solid axes of the boxes shows the direction of the profile, which were collapsed along the dotted axes. The profiles were extracted from the raw red image for the four extraction regions shown. A typical stellar profile (FWHM ~4 pixels) is overlaid in red, which is less extended than any of the filament profiles ($\geq 7$ pixels). The x-axis is in units of arcsec (pixels are 0.025 arcsec wide), the y-axis is intensity in arbitrary units.

# Supplementary Table 1 – Log of observations

| Pointing | RA | Dec | Date | Filter | Exposure(s) | pa(deg) |
|---|---|---|---|---|---|---|
| NGC1275-SW | 3 19 45.3 | 41 30 40.0 | 2006-07-29 | F435W | 815.0 | 260.3 |
| NGC1275-SW | 3 19 45.6 | 41 30 40.7 | 2006-07-29 | F435W | 815.0 | 260.3 |
| NGC1275-SW | 3 19 45.8 | 41 30 41.5 | 2006-07-29 | F435W | 815.0 | 260.3 |
| NGC1275-SW | 3 19 45.3 | 41 30 40.0 | 2006-07-30 | F435W | 824.0 | 260.3 |
| NGC1275-SW | 3 19 45.6 | 41 30 40.7 | 2006-07-30 | F435W | 824.0 | 260.3 |
| NGC1275-SW | 3 19 45.8 | 41 30 41.5 | 2006-07-30 | F435W | 824.0 | 260.3 |
| NGC1275-SE | 3 19 50.5 | 41 30 40.0 | 2006-08-04 | F435W | 815.0 | 257.9 |
| NGC1275-SE | 3 19 50.8 | 41 30 40.9 | 2006-08-04 | F435W | 815.0 | 257.9 |
| NGC1275-SE | 3 19 51.0 | 41 30 41.7 | 2006-08-04 | F435W | 815.0 | 257.9 |
| NGC1275-SE | 3 19 50.5 | 41 30 40.0 | 2006-08-05 | F435W | 824.0 | 257.9 |
| NGC1275-SE | 3 19 50.8 | 41 30 40.9 | 2006-08-05 | F435W | 824.0 | 257.9 |
| NGC1275-SE | 3 19 51.0 | 41 30 41.7 | 2006-08-05 | F435W | 824.0 | 257.9 |
| NGC1275-NW | 3 19 45.3 | 41 31 60.0 | 2006-08-08 | F550M | 806.0 | 256.4 |
| NGC1275-NW | 3 19 45.3 | 41 31 60.0 | 2006-08-08 | F550M | 806.0 | 256.4 |
| NGC1275-NW | 3 19 45.6 | 41 32 0.9 | 2006-08-08 | F550M | 806.0 | 256.4 |
| NGC1275-NW | 3 19 45.6 | 41 32 0.9 | 2006-08-08 | F550M | 806.0 | 256.4 |
| NGC1275-NW | 3 19 45.8 | 41 32 1.9 | 2006-08-08 | F550M | 806.0 | 256.4 |
| NGC1275-NW | 3 19 45.8 | 41 32 1.9 | 2006-08-08 | F550M | 806.0 | 256.4 |
| NGC1275-SW | 3 19 45.3 | 41 30 40.0 | 2006-07-29 | F550M | 806.0 | 260.3 |
| NGC1275-SW | 3 19 45.3 | 41 30 40.0 | 2006-07-29 | F550M | 806.0 | 260.3 |
| NGC1275-SW | 3 19 45.6 | 41 30 40.7 | 2006-07-29 | F550M | 806.0 | 260.3 |
| NGC1275-SW | 3 19 45.6 | 41 30 40.7 | 2006-07-29 | F550M | 806.0 | 260.3 |
| NGC1275-SW | 3 19 45.8 | 41 30 41.5 | 2006-07-29 | F550M | 806.0 | 260.3 |
| NGC1275-SW | 3 19 45.8 | 41 30 41.5 | 2006-07-29 | F550M | 806.0 | 260.3 |
| NGC1275-SW | 3 19 45.3 | 41 30 40.0 | 2006-07-29 | F550M | 813.0 | 260.3 |
| NGC1275-SW | 3 19 45.3 | 41 30 40.0 | 2006-07-29 | F550M | 813.0 | 260.3 |
| NGC1275-SW | 3 19 45.6 | 41 30 40.7 | 2006-07-29 | F550M | 813.0 | 260.3 |
| NGC1275-SW | 3 19 45.6 | 41 30 40.7 | 2006-07-29 | F550M | 813.0 | 260.3 |
| NGC1275-SW | 3 19 45.8 | 41 30 41.5 | 2006-07-29 | F550M | 813.0 | 260.3 |
| NGC1275-SW | 3 19 45.8 | 41 30 41.5 | 2006-07-29 | F550M | 813.0 | 260.3 |
| NGC1275-SE | 3 19 50.5 | 41 30 40.0 | 2006-08-04 | F550M | 806.0 | 257.9 |
| NGC1275-SE | 3 19 50.5 | 41 30 40.0 | 2006-08-04 | F550M | 806.0 | 257.9 |
| NGC1275-SE | 3 19 50.8 | 41 30 40.9 | 2006-08-04 | F550M | 806.0 | 257.9 |
| NGC1275-SE | 3 19 50.8 | 41 30 40.9 | 2006-08-04 | F550M | 806.0 | 257.9 |
| NGC1275-SE | 3 19 51.0 | 41 30 41.7 | 2006-08-04 | F550M | 806.0 | 257.9 |
| NGC1275-SE | 3 19 51.0 | 41 30 41.7 | 2006-08-04 | F550M | 806.0 | 257.9 |
| NGC1275-SE | 3 19 50.5 | 41 30 40.0 | 2006-08-04 | F550M | 813.0 | 257.9 |
| NGC1275-SE | 3 19 50.5 | 41 30 40.0 | 2006-08-04 | F550M | 813.0 | 257.9 |
| NGC1275-SE | 3 19 50.8 | 41 30 40.9 | 2006-08-04 | F550M | 813.0 | 257.9 |
| NGC1275-SE | 3 19 50.8 | 41 30 40.9 | 2006-08-04 | F550M | 813.0 | 257.9 |
| NGC1275-SE | 3 19 51.0 | 41 30 41.7 | 2006-08-05 | F550M | 813.0 | 257.9 |
| NGC1275-SE | 3 19 51.0 | 41 30 41.7 | 2006-08-05 | F550M | 813.0 | 257.9 |
| NGC1275-NW | 3 19 45.3 | 41 31 60.0 | 2006-08-08 | F625W | 827.0 | 256.4 |
| NGC1275-NW | 3 19 45.6 | 41 32 0.9 | 2006-08-08 | F625W | 827.0 | 256.4 |
| NGC1275-NW | 3 19 45.8 | 41 32 1.9 | 2006-08-08 | F625W | 827.0 | 256.4 |
| NGC1275-SW | 3 19 45.3 | 41 30 40.0 | 2006-07-29 | F625W | 827.0 | 260.3 |
| NGC1275-SW | 3 19 45.6 | 41 30 40.7 | 2006-07-29 | F625W | 827.0 | 260.3 |
| NGC1275-SW | 3 19 45.8 | 41 30 41.5 | 2006-07-29 | F625W | 827.0 | 260.3 |
| NGC1275-SW | 3 19 45.3 | 41 30 40.0 | 2006-07-29 | F625W | 827.0 | 260.3 |
| NGC1275-SW | 3 19 45.6 | 41 30 40.7 | 2006-07-29 | F625W | 827.0 | 260.3 |
| NGC1275-SW | 3 19 45.8 | 41 30 41.5 | 2006-07-30 | F625W | 827.0 | 260.3 |
| NGC1275-SE | 3 19 50.5 | 41 30 40.0 | 2006-08-04 | F625W | 827.0 | 257.9 |
| NGC1275-SE | 3 19 50.8 | 41 30 40.9 | 2006-08-04 | F625W | 827.0 | 257.9 |
| NGC1275-SE | 3 19 51.0 | 41 30 41.7 | 2006-08-04 | F625W | 827.0 | 257.9 |
| NGC1275-SE | 3 19 50.5 | 41 30 40.0 | 2006-08-05 | F625W | 827.0 | 257.9 |
| NGC1275-SE | 3 19 50.8 | 41 30 40.9 | 2006-08-05 | F625W | 827.0 | 257.9 |
| NGC1275-SE | 3 19 51.0 | 41 30 41.7 | 2006-08-05 | F625W | 827.0 | 257.9 |